# Interfacial Ionic "Liquids": Connecting Static and Dynamic Structures

Ahmet Uysal,[1,*] Hua Zhou,[2] Guang Feng,[3,*] Sang Soo Lee,[1] Song Li,[3] Peter T. Cummings,[3] Pasquale F. Fulvio,[4,#] Sheng Dai,[4] John K. McDonough,[5] Yury Gogotsi[5] and Paul Fenter[1,*]

[1] Chemical Science and Engineering Division, Argonne National Laboratory, Argonne, IL 60439

[2] Advanced Photon Source, Argonne National Laboratory, Argonne, IL 60439

[3] Department of Chemical and Biomolecular Engineering, Vanderbilt University, Nashville, TN 37235

[4] Chemical Sciences Division, Oak Ridge National Laboratory, Oak Ridge, TN 37831

[5] Department of Materials Science and Engineering & A.J. Drexel Nanomaterials Institute, Drexel University, Philadelphia, PA 19104

E-mail: ahmet@anl.gov (A.U.); gfeng@hust.edu.cn (G.F.); pfenter@anl.gov (P.F.)

It is well-known that room temperature ionic liquids (RTILs) often adopt a charge-separated layered structure, i.e., with alternating cation- and anion-rich layers, at electrified interfaces. However, the dynamic response of the layered structure to temporal variations in applied potential is not well understood. We used *in situ*, real-time X-ray reflectivity (XR) to study the potential-dependent electric double layer (EDL) structure of an imidazolium-based RTIL on charged epitaxial graphene during potential cycling as a function of temperature. The results suggest that the graphene-RTIL interfacial structure is bistable in which the EDL structure at any intermediate potential can be described by the combination of two extreme-potential structures whose proportions vary depending on the polarity and magnitude of the applied potential. This picture is supported by the EDL structures obtained by fully atomistic molecular dynamics (MD) simulations at various static potentials. The potential-driven transition between the two structures is characterized by an increasing width but with an approximately fixed hysteresis magnitude as a function of temperature. The results are consistent with the coexistence of distinct anion and cation adsorbed structures separated by an energy barrier (~0.15 eV).

# I. INTRODUCTION

The continuous development of more powerful electronic devices and our growing dependence on them have increased the demands from advanced energy storage components. For instance, future generation electric energy storage devices should have longer cycle life, higher power density, and shorter charging times along with the high energy density [1, 2]. Electric double layer capacitors (EDLCs) or supercapacitors in general, are promising candidates to satisfy these goals - alone or in tandem with rechargeable batteries. EDLCs made with highly conductive low dimensional carbon electrodes, such as graphene, and room temperature ionic liquids (RTILs) can lead to outstanding properties [2-4].

RTILs, electrolyte liquids consisting of only cations and anions without any solvent, behave very differently from more widely studied dilute electrolytes at charged interfaces [5]. The distinct features of the electric double layer (EDL) structure for RTILs include the short range ion-ion interactions, the finite size of the ions and the irregularities in the ion shape. These features are not well described with classical mean-field theories, such as Gouy-Chapman theory. Modifications to this theory to include the finite size of ions, successfully described some important properties of RTILs [5, 6], and non-mean field theories suggested explanation to RTILs' unexpectedly high double layer capacitance [7, 8]. However, we are still far from having a realistic theoretical picture of the RTIL/electrode interface.

Recent experimental and computational studies show that RTILs can exhibit charge-separated (cation-anion) layered structures at charged interfaces [5, 6, 9-15]. These results are not totally unexpected, considering that many liquids (including liquid metals [16], dielectric liquids [17], and even water [18]) are known to exhibit layering at the liquid/solid interface. However, the effects of ion-separated layers on the dynamical interfacial response of RTILs are not well understood. In a recent review, this problem is stated as "one of the possible central subjects for future studies" in RTIL research [5]. In particular, recent experimental observations of potential dependent hysteresis of the EDL structure [19-25] and its ultra-slow response to potential steps [25-29] remain unexplained by any theoretical foundation. Both of these phenomena, which are unexpected in liquids, may elucidate the intrinsic properties of RTILs. However, most theoretical and computational methods do not cover the relevant time scales revealed by experimental studies [19-29] and the existing experiments are not conclusive as to the origin of this behavior.

There are multiple distinct timescales that may be involved in RTIL dynamics. The sub-nanosecond time scale, which primarily reflects the diffusion of molecules, can be studied by experimental techniques such as nuclear magnetic resonance (NMR) [30] and quasi elastic neutron scattering (QENS) [31] and is easily accessible to computational studies, such as molecular dynamics (MD) simulations [10, 31-33] and time-dependent density functional theory calculations [34, 35]. Timescales in the millisecond regime cover the main capacitive processes due to the EDL and can be understood in terms of an RC constant of the system ($10\mu F \times 1k\Omega = 10$ ms, for a typical EDL capacitor with 1 cm$^2$ surface area) and have been studied by electrochemical impedance spectroscopy (EIS). Some recent computational efforts have explored the RTIL response at these time scales with equivalent circuit models [32]. Timescales of seconds and longer are unexpected for a liquid. No physical parameter of an RTIL (e.g., diffusion constant) can explain such long time scales.

In one of the first systematic EIS studies of RTILs, Lockett et al. [19] reported hysteresis in differential capacitance (DC) measurements and a slow process with a time constant of the order of seconds. Both observations were attributed to the specific adsorption of imidazolium cations by the electrode. Because these cations are large organic molecules, the authors suggested that the process is analogous to the hysteresis observed with organic electrolytes in aqueous solutions. Indeed, there is a vast literature about the potential-driven phase transitions at electrochemical interfaces [36] and they are being revisited to develop a better understanding of RTILs [37]. However, as Lockett et al. stated in a later study [20], the hysteresis observed in aqueous solutions are mainly related to desorption of water from the metal electrodes and cannot be directly compared to RTILs which have no solvent.

Various explanations for the hysteresis have been offered. It was suggested that the observed hysteresis could be a result of the slow response of the RTIL itself. In this scenario, it would be expected that slower scan rates would lead to reduced hysteresis [20, 21]. Effects of the electrodes were considered as another possible cause. Since most of the electrodes in those studies were metals, whose surfaces are known to exhibit charge density dependent surface reconstructions [36], surface reconstructions have been suggested as the origin of the slow processes [26]. Pseudocapacitive processes are another possible reason for slow processes and the hysteresis [38]. These discussions illustrate the need for experimental probes, such as X-ray scattering and surface specific spectroscopies, that can distinguish between the effects caused by the RTIL from those due to the electrode surface through direct measurements.

Sum frequency generation (SFG) studies under static potentials with long waiting times between measurements ruled out the possibility that the observed hysteresis is only a kinetic effect [22, 23]. More interestingly, recent surface-enhanced infrared absorption spectroscopy (SERIAS) [24] and our X-ray reflectivity (XR) [15] measurements showed that the relation between the slow response and the structural hysteresis is counterintuitive, i.e. the hysteresis becomes more pronounced at slower scan rates.

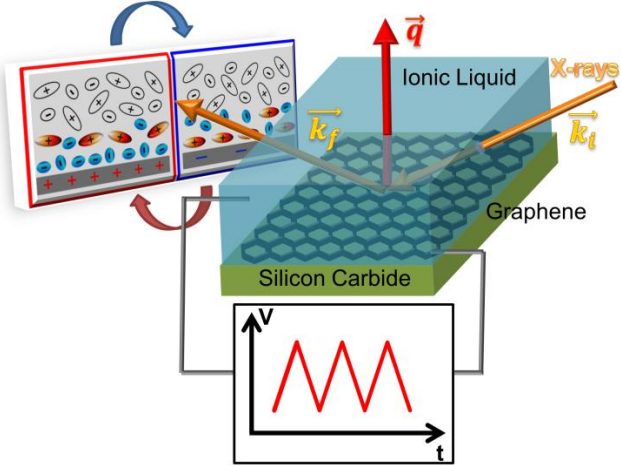

**Figure 1**. A schematic representation of the XR scattering geometry from the EG/RTIL interface under electrochemical control and the simplified models of the RTIL interfacial structures exhibiting a bistable character as explained in the text (not drawn to scale).

XR is a direct experimental probe of the *in situ* solid/liquid interfacial *structure* [12-17, 39, 40]. There are few XR studies of RTIL/charged solid interfaces [12-15], only two of which were performed with controlled electrode potentials [13, 15]. Our recent study of an imidazolium RTIL on epitaxial graphene (EG) electrodes revealed two primary observations: the RTIL structures at extreme potentials (i.e., -0.4 V and 1.0 V) are described by extended interfacial RTIL distributions with layered, charge-separated, profiles that can be distinguished by the surface adsorbed ion (i.e., cation or anion); and the potential-controlled hysteresis and slow dynamic response to potential steps are associated with the interfacial RTIL structure [15] (i.e., the XR data show that these behaviors are not due to potential-dependent reconstructions of the electrode surface).

Here, we present *real-time* XR observations of RTIL interfacial structures at EG during potential cycling to investigate the dynamic response of the interfacial RTIL and to develop a new conceptual model of their potential-controlled structures and hysteresis. These XR experiments, performed at multiple temperatures, are integrated with fully atomistic molecular dynamics (MD) simulations. Our results suggest that the intrinsically layered EDL structure has significant effects on its response to changes in potential, and its dynamical response is explained by a bistable model. The model provides an estimate for the size of the energy barrier for the transition between two extreme-potential structures.

## II. METHODS

We use RTIL 1-Methyl-3-nonylimidazolium Bis-(trifluoromethanesulfonyl)imide ($[C_9mim^+]$ $[Tf_2N^-]$); details of its synthesis were described previously [15]. The neat RTIL was annealed in a vacuum furnace at 80°C for 48 h before the experiment to remove any volatile impurities, including water. Few-layer EG was grown on 6H silicon carbide (SiC) substrates (Cree) under argon environment (600-900 mbar) in an MTI GSL1700X tube furnace at 1550 °C similar to a method described in Ref. [41]. A sample cell, using transmission geometry with electrochemical and temperature control, was used for *in situ* XR experiments (as shown schematically in Fig. 1) [42]. The EG working electrode and Pt pseudo-reference and counter electrodes were controlled by a Gamry Reference-600 potentiostat. Measurements were limited to potentials between -0.4 V and 1.0 V, in the middle of the ~2.5 V electrochemical window where the processes are capacitive. Synchrotron XR experiments were conducted at 6-IDB, 12-IDD and 33-IDD of Advanced Photon Source at Argonne National Laboratory. The monochromatic X-ray photon energy was 20keV ($\lambda$=0.62 Å). The incident beam was focused vertically (80 μm) with a horizontal size of ~1mm. Specular XR data, the ratio of the intensity of the reflected X-ray signal from the RTIL/EG interface to the incoming X-ray intensity, were recorded as a function of vertical momentum transfer **q**, where **q** = $\mathbf{k}_f$-$\mathbf{k}_i$ = $(4\pi/\lambda)(\sin2\theta/2))$ [Fig. 1]. MD simulations were performed in the NVT ensemble using a customized MD code based on the Gromacs 3.3 software [43]. A detailed description of these simulations is available elsewhere [15]; however, the current simulations were run for longer times, with 12 ns for annealing, 20 ns for equilibrium and 30 ns for production.

## III. RESULTS
### A. Real-time XR during potential cycling

Time-resolved XR measurements, R(t, $\mathbf{q}_o$), were obtained during potential cycling, V(t), at selected $\mathbf{q}_o$ from 0.15Å$^{-1}$ to 0.35Å$^{-1}$. Static potential experiments [15] show that the signal in this region is sensitive to the RTIL structure. These data are used to construct a time-dependent XR signal contour, R(**q**, t) [Fig. 2a], that can be interpreted as real-time potential-dependent reflectivity curves, R(**q**, V(t)) [Fig. 2b, symbols]. These data are directly sensitive to the *dynamic* structural evolution of the interfacial RTIL (including the presence of any charge separated anion/cation layers) obtained with ~1 sec time resolution.

The data in Fig. 2b reveal that the XR signal changes monotonically with applied potential at all measured **q** and the position of the reflectivity minimum does not change during potential cycling. Our previous static potential measurements of the same system [15] show that while the position of this minimum is determined by the distance between the charge separated layers in the diffuse part of the EDL, its depth is sensitive to the characteristics of the bound part of the EDL (usually named as Helmholtz or Stern layer). This suggests that as the potential is cycled the layered structure does not change but the charged species swap their positions. The reflectivity at any intermediate potential, R(**q**, V), may be described by a linear combination of those observed at extreme potentials (-0.4 V and 1.0 V). That is, we can write [44]:

$$R(\mathbf{q}, V) = n(V) R_N(\mathbf{q}) + (1 - n(V)) R_P(\mathbf{q}) \quad (1)$$

Here $R_N$ and $R_P$ are the reflectivity data at the extreme negative (-0.4 V) and positive (1.0 V) potentials, respectively. The calculated curves (lines, Fig. 2b) for intermediate potential structures derived using a weighting factor, $n(V)$ in Eq. 1, where $n(V)$ and 1- $n(V)$ are the weights of the reflectivity signals at the extreme negative and positive potentials, respectively. Eq. 1 shows very good agreement with the data for all **q** (symbols, Fig. 2b). R(**q**, V) curves for the cathodic scan, from -0.4 V to 1.0V are shown (similar curves were also obtained from Fig. 2a for the anodic scan, 1.0V to -0.4V, not shown). The $n(V)$ values obtained for one full CV cycle (averaged over 3 repeated measurements) [black circles, Fig. 2c] reveals two observations: the potential-controlled RTIL structure exhibits a broad transition as a function of applied potential, and also shows structural hysteresis (for V>0.4 V). The lack of visible hysteresis for V<0.3 V is consistent with the potential-dependent saturation that starts around the same potential. The saturation limit at high positive potentials is out of the accessible electrochemical window. These results are consistent with the hysteresis we observed [15] in the XR signal at a single scattering condition.

Fig. 2b and Eq. 1 have multiple implications concerning the RTIL structure during potential cycling. First of all, they show that the layered interfacial structure is not disturbed during the cycling because the position of the minimum does not change significantly. Therefore, we conclude that the layered interfacial structure is relatively robust as a function of applied potential. More significantly, since the XR data represent the laterally averaged signal over a macroscopic area (~1 mm$^2$), our experimental results suggest that the potential dependence of the EDL structure is controlled by a mixture of the two extreme-potential RTIL structures, $\rho_N(z)$ and $\rho_P(z)$, with fractional coverages of each structure at potential V of $n(V)$ and 1-$n(V)$ respectively, where z-dimension is normal to the graphene electrode. Thus, the EDL structure described by the laterally averaged vertical electron density profile (EDP) as a function of the applied potential can be written as:

$$\rho(V, z) = n(V) \rho_N(z) + (1 - n(V)) \rho_P(z) \quad (2)$$

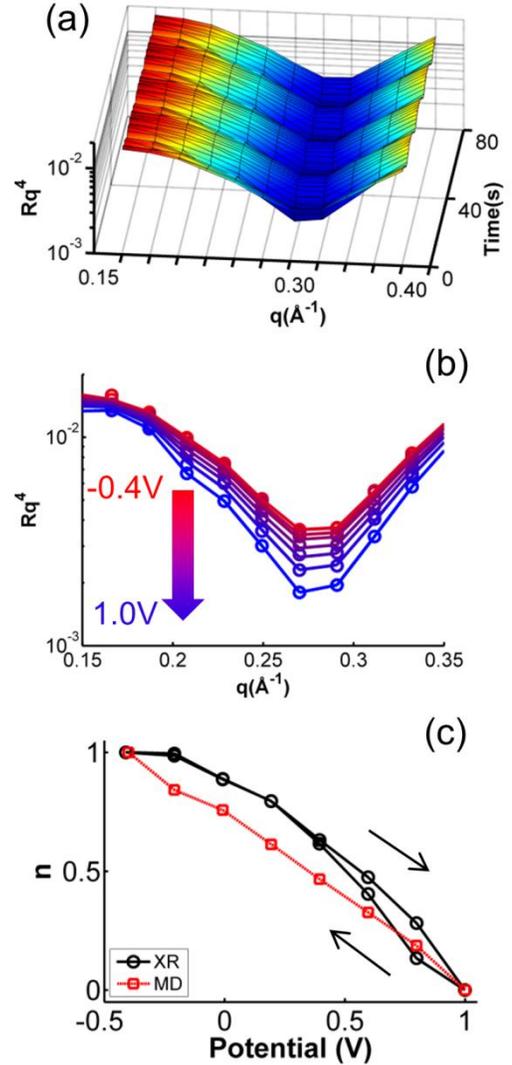

**Figure 2**. (a) Real-time XR signals, R(**q**,t), shown as contours, reveal changes in the reflectivity signal during CV cycling at 100mV/s. (b) Potential-dependent XR data (symbols) during the cathodic scan (Error bars are smaller than the symbols). The lines show the best fits according to Eq. 1. (c) Potential-dependent weighting factors derived from the best fits to the XR data show hysteresis (black circles). The same weighting factors are also determined from EDPs calculated from static MD simulations (Figure 3a) for comparison (red squares).

That is, the interfacial structure locally exhibits bistable behavior, as a mixture of cation and anion-terminated regions. Therefore, both the width of the potential-driven transition and the potential-dependent hysteresis are controlled by the relative coverages of the two phases, $n(V)$ and 1- $n(V)$. The presence of hysteresis suggests that the transition between these two phases is controlled by an energy barrier, presumably derived from the steric energy cost associated with reorganizing the extended interfacial RTIL structure.

Since the hysteresis becomes more pronounced at slower scan rates [15, 24], $n(V)$ is expected to depend on the scan rate. We measured the time-dependent XR curves at 100 mV/s because the time necessary for similar measurements at 5 mV/s is prohibitive (~20 times longer). However, we checked a few fixed scattering conditions at 5mV/s (similar to the one reported in Ref. 15) and we observed that ΔV is ~0.15±0.03 V. Therefore, we conclude that $n(V)$ values obtained from 100 mV/s scans are representative of a wide range of conditions.

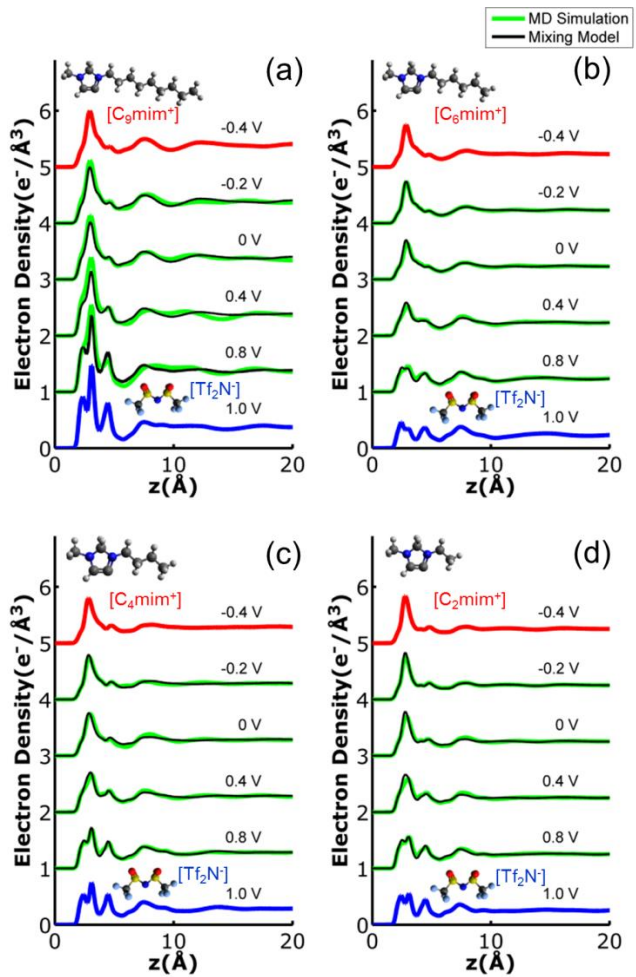

**Figure 3**. Potential-dependent EDPs calculated from MD simulations at various static potentials (red, green and blue lines). The Stern layer is cation and anion rich at -0.4 V and 1.0 V respectively. The black lines at intermediate potentials are calculated according to Eq. 2, in good agreement with MD simulations.

**B. MD Simulations**

Further insights are explored by comparing these results to fully atomistic MD simulations of this and related RTIL systems (with cation chain lengths ranging from 2 to 9 carbons) at various *static* potentials. The goal here is to compare the predictions of Eq. 2 to the results obtained from MD simulations. The simulations have been equilibrated by annealing at elevated temperatures at each potential, and therefore are only expected to reflect the equilibrium variation of the RTIL structure vs. potential, without any history effects.

The interfacial structures were simulated at two extreme potentials (-0.4 and 1 V; solid blue and red lines, respectively, Fig. 3a) and four intermediate potentials (-0.2, 0, 0.4 and 0.8 V; green solid lines, Fig. 3a). The results show the formation of distinct (and denser) cation and anion layers adsorbed at the negatively and positively-charged EG surface, respectively. Interestingly, these adsorbed layers are observed even at the intermediate potentials with minimal changes in their adsorption geometry and conformation. Only the proportions between the adsorbed coverages of anion and cation species change. The latter result is consistent with the observations from XR. To quantify the similarity between MD and XR results, we compared the MD-simulated EDPs at the intermediate potentials with those predicted from Eq. 2 by using the *experimentally*-derived $n(V)$ [Fig. 2c] (using the average $n(V)$ values obtained during anodic and cathodic scans). The result shows good agreement between the predicted and simulated profiles (black and green lines, respectively), especially for the portion of the density profile closest to the EG interface where the ion layering is strongest. The same approach was applied for the individual profiles of cations or anions, and the results (not shown) yielded essentially the same agreement. The generality of these observations was further explored computationally for three other RTILs with different cations ([$C_6mim^+$], [$C_4mim^+$] and [$C_2mim^+$]), successfully reproducing the potential dependent MD simulation results (Fig. 3b, c and d respectively).

We have also determined the weighting factors, $n(V)$, directly from the MD simulated structures (red squares, Fig. 2c) for the RTIL with the [$C_9mim^+$] cation. Here, the MD-derived $n(V)$ (from Eq. 2) show a broad transition, similar to that observed experimentally, but vary more linearly with the potential. That is, the average RTIL structure at intermediate potentials (under equilibrium conditions) is well-described by a potential-controlled mixture of the extreme potential structures, as opposed to a continuous variation of the RTIL interfacial structure (e.g., sorption geometry or conformation of the molecules) as a function of potential. This non-trivial result provides direct confirmation of the RTIL bistability inferred from the real-time XR data (Fig. 2).

While there are numerous similarities between the real-time potential-dependent RTIL response as observed by XR and the equilibrium response obtained by MD, there are also two differences. The most significant difference is the evident role of hysteresis in the XR, which could not be addressed in the static MD simulations with fixed surface charge. The second difference is that the transition, as observed by XR, is characterized by an error-function-like transition which is distinct from the linear response observed from MD simulations. The fixed charged calculations in our MD simulations may be the main reason for the discrepancy in n(V) values obtained by experiment and MD-to-MD fits in Fig 2c. Indeed, Merlet et.al. [10] report that the potential-dependent variation of differential capacitance gets flatter if the fixed potential approach and polarized force fields are not used, consistent with a linear variation of $n$(V), similar to the ones we obtain in our MD simulations. Qualitatively, this suggests that the implied differential capacitance as observed by XR might be described by a bell-shaped curve whose width is narrower than that obtained by MD simulations [45].

The suggested bistable EDL model requires the coexistence of oppositely polarized RTIL structures. Such a structure might be stabilized at an electrode surface having a non-uniform charge distribution (i.e., at a given applied potential; unlike the constant charge model used in the present MD simulations). Indeed, this suggestion is consistent with predictions of recent non-mean-field theories [7, 8] and computational studies under controlled electrode potentials [10].

### C. Temperature dependent measurements

The generality of these results can be tested by making observations of the real-time reflectivity signals during CV cycling as a function of temperature (at T = 25, 60 and 100 °C; Fig. 4a). The XR data were collected at $q$=0.3Å$^{-1}$ at a scan rate of 100 mV/s. The time dependent data [Fig. 4a], averaged over 3 cycles, reveal the temperature dependence of the potential-dependent hysteresis curves (symbols, Fig. 4b). There are visible changes to the temporal variation of the reflectivity signal, with an inverted parabolic response at 25 °C and a more saw-tooth shape at 100 °C (Fig. 4a). Nevertheless, it is apparent from these data (when plotted as a function of applied potential) that hysteresis is observed over the full range of temperatures (Fig. 4b).

A deep understanding of these observations relies on developing a conceptual model for the potential dependent RTIL structure. The hysteresis that we observed appears to be a characteristic of systems that exhibits a first order phase transition, consistent with a recent computational study [10]. However, the results also reveal that the transition is very broad in applied potential (> 1V), which is unexpected for a first order transition. Since the fundamental nature of this transition remains unclear, we characterize the potential- and temperature-dependent behavior based on a simple phenomenological model.

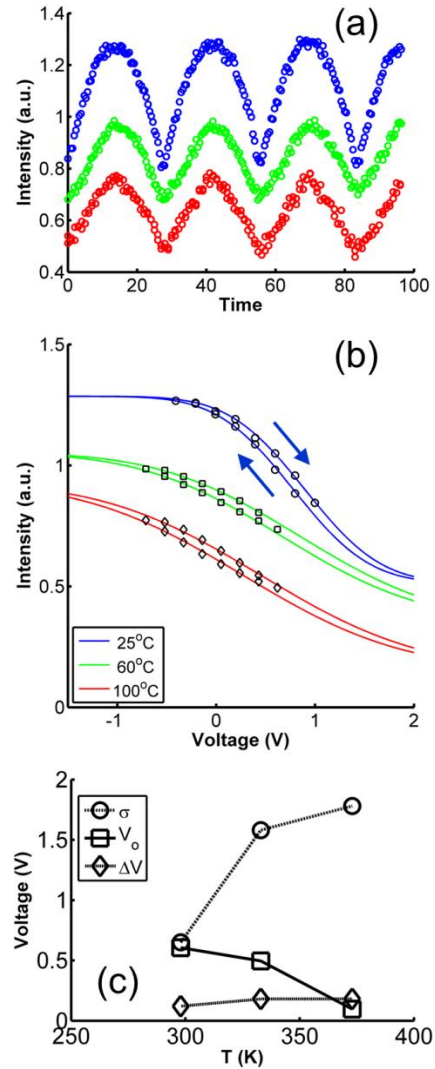

**Figure 4**. (a) Time-dependent XR signal at various temperatures ($q$ = 0.3 Å$^{-1}$, CV cycling rate = 100 mV/s). (b) Potential-dependent XR data (symbols) derived from the time-dependent XR data in (a) (arrows indicate the cycling direction). The fit lines are calculated according to the bistable model described in text. (The 60°C and 100°C data are shifted by 0.5 and 0.4 units respectively in both (a) and (b)) (c) Temperature dependence of the Gaussian parameters describing the transition are shown in (c), including the transition width ($\sigma_V$), the center of the hysteresis curve ($V_0$), and the hysteresis magnitude as indicated by the separation ($\Delta V$) between the cathodic and anodic scan curves.

It has been shown that the hysteretic response of any bistable system (i.e., with two stable states separated by an energy barrier) can be described as a linear combination of contributions from individual domains (without any restriction on the physical shape or size of domains) [46, 47]. Here, the two stable states correspond to the cation- and anion-adsorbed EDPs observed at the extreme potentials (Fig. 3a, blue and red curves, respectively). If we were to consider the response of a single bistable "domain", an energetic barrier between the two states would lead to a rectangular hysteresis loop in a state vs.

potential plot, with a hysteresis magnitude, ΔV, centered at $V_0$ (that is, with transition potentials at $V_0 \pm \Delta V/2$ for the two sweep directions). The finite width ($\sigma_V$) of the observed transition can be understood as a result of a number of factors, including charge inhomogeneities within the graphene layer [10], thermal fluctuations or in-plane ion-ion interactions. The switching would then occur over a range of potentials, characterized by a voltage width, $\sigma_V$, distributed around the intrinsic transition voltages, $V_0 \pm \Delta V/2$. Our results are consistent with a Gaussian distribution in the transition voltage, so that the observed hysteresis loop can be characterized by two error function profiles that are separated in voltage by ΔV. The results of this analysis [Fig. 4b, solid lines] allow us to characterize the temperature dependent response of this system [Fig. 4c].

These results show that the hysteresis magnitude, ΔV ≈ 0.15 V, is largely independent of temperature within the temperature range studied. Within the context of the phenomenological model of bistable systems, the hysteresis parameter characterizes an effective energy barrier that separates the two coexisting states (0.15 eV, or ~9 $k_BT_g$, where $T_g$=193 K is the glass transition temperature [48]). It is interesting to note that dielectric spectroscopy studies have inferred glass relaxation energy barriers of this magnitude for comparable RTILs [49]. The presence of this larger barrier may explain the persistence of hysteresis at elevated temperatures. It also provides further evidence for the inherent bistability of the RTIL interfacial structure. The largest sensitivity of the RTIL response is found in the transition width, $\sigma_V$, which increases substantially with increasing temperature (by nearly a factor of 4). Therefore, the effective heterogeneity of RTIL-graphene interactions apparently increases with temperature. Finally, we also observe a shift of the average transition voltage, $V_0$, towards more negative potentials with the increasing temperature. This may be more specific to the RTIL used in this study. As we have shown previously [14], due to the π-π interactions between the imidazolium ring and the EG electrode, the first adsorbed layer has more cations than anions even without applied potential, and applying increasingly negative potentials does not affect the cation-adsorbed structure significantly. Apparently, increasing temperatures weakens the cation-graphene interaction so that more negative potentials are needed to stabilize cation adsorption.

## IV. DISCUSSION

The present results provide the first direct and real-time observations of an RTIL/electrode interfacial structure during real-time potential scanning. Our results, as characterized by a simple model, suggest that the observed behavior derives from the combination of an inherently bistable RTIL structure and a heterogeneity of the RTIL-graphene interactions. According to this picture, the change in the observed EDL structure as a function of potential is defined by the fractional ratio of the areas covered by cation and anion adsorbed regions.

These results provide the first experimental observations related to the recent computational and theoretical predictions about the potential-dependent EDL structure of RTILs. For example two recent coarse-grain MD simulations, one with constant electrode potential [10] and the other with constant surface charge [50], predicted a potential-dependent order-disorder phase transition in the Stern layer. The latter study also predicted a multilayer-monolayer phase transition in the z-direction perpendicular to the interface. The phenomenological model that we use to characterize our results is expressed as the response of "domains" of cation and anion adsorbed regions. However, the applicability of this model is independent of the size and the shape of any such "domains" and does not require the in-plane ordering of ions. Experimental evidence for any such laterally extended structures with X-ray scattering can be obtained from non-specular (in-plane) scattering from the interface. Those experiments will be subject of further studies. On the other hand, both XR data and the MD simulations presented here do not show any evidence for disturbance of the layered EDL structure as a function of potential (beyond that due to the observed bistability). Therefore, within the potential window that we studied, we did not find any evidence of a multilayer-monolayer transition.

We have previously found that the extreme-potential RTIL structures are characterized by alternating charged layers [15]. Consequently, the potential-dependent switching between these structures cannot be explained by the adsorption of a single cation or anion layer, or by a conformational change in the first RTIL layer. Instead, the switching inherently requires the simultaneous reorganization of cations and anions within an extended number of layers, with coordinated motions of both the anions and cations in opposite directions. This suggests that the observed $9k_BT_g$ energy barrier that separates the two structures may be a steric barrier between two structures which is independent of temperature. It is interesting that earlier EIS studies have suggested a temperature-independent energy barrier for slow capacitive processes at RTIL/gold interface [26]. A recent publication has also proposed that in-plane heterogeneities in the Stern layer can be used to explain the slow processes theoretically [51]. However, in the latter study, the authors assumed that the different orientations of ions at the interface are responsible for the heterogeneity. In contrast, our experiments suggest that distinct anion/cation layered structures (characterized by the first adsorbed RTIL layer) is the origin of the bistability at the interface. As discussed in the introduction, the observed slow processes and hysteresis are likely distinct, but they may have similar origins, i.e. the steric energy barrier may be causing both the slow response and the hysteresis. A solution to this

"mystery", as Fedorov and Kornyshev referred [5], is possible only with more experimental studies on the structural dynamics at the interface with different types of RTILs and electrodes to identify system specific effects.

An important difference in our studies with respect to previous studies [19-29] is in the use of graphene as a substrate. The robust structure of graphene (without any known potential-controlled surface reconstructions) immediately suggests that these slow relaxations originate from the structural changes in the interfacial RTIL structure alone. However, the strong π-π interaction between the imidazolium ring of the cation and the graphene, and the semimetal nature of graphene [52] may be responsible for some of the observed effects. Experiments with different cations will provide further insights to this issue.

An understanding of the role of impurities in the RTIL properties must be considered, especially water [5, 53]. The present measurements were performed using vacuum dried RTILs which are reported to have ~100 ppm water even after the treatment [53]. Therefore, it is reasonable to ask if a water impurity might have any effect on our results. Firstly, we repeated some of the measurements with RTILs that were exposed to the atmosphere for a long time, and the results showed no significant difference in the observed behavior (not shown). Secondly and more importantly, the bistable EDL behavior that we observed based on the XR data, was reproduced by the potential dependent MD simulations (Fig. 3), which did not include any impurities.

What is the nature of the interfacial RTIL? There are several studies in the literature [54, 55] that use the term "solid-like" to describe interfacial RTILs. Such an organized structure would lead to significant restrictions on the ion movements; but it is also apparent that the RTIL structure is controlled by the applied potential, and therefore is not well-described as a 'solid'. On the other hand, the vast differences between the ~ns response time for diffusion and the >1 s time constant that we observe at the interface cannot be explained only by the smaller diffusion constant of fluids at solid interfaces compared to their bulk values [56]. Our experiments reveal that the RTIL structure consists of inherently layered (and charge separated) structure at all applied potentials (but with coexistence of cation and anion adsorbed regions). Therefore, we believe that the properties are best described as an "interfacial ionic liquid", stressing the significantly different structure and dynamics compared to that in the bulk.

## V. SUMMARY

Direct observations of the structural dynamics at the RTIL/EG interfaces during real-time CV cycling reveal that the previously observed hysteresis in these systems is due to a potential-dependent switching between two extreme potential structures, characterized by cation- and anion-adsorbed phases. These results provide a direct connection between the previously identified static EDL structures and those achieved during dynamic conditions. A simple phenomenological model explains the observed hysteresis in terms of a bistable structure stablized by an energy barrier of ~9 $k_B T_g$, although a more complete picture of the interactions leading to this behavior is needed. We propose that the presence of charge-separated and layered interfacial RTILs is central to this behavior.

## ACKNOWLEDGMENTS

We thank Tim T. Fister, Francesco Bellucci and Nouamane Laanait for technical and intellectual support at various stages of the experiments. We also thank Brian Skinner for valuable discussions. This work was supported as part of the Fluid Interface Reactions, Structures and Transport (FIRST) Center, an Energy Frontier Research Center funded by the U.S. Department of Energy, Office of Science, Office of Basic Energy Sciences. Use of the beamlines 6ID, 12ID and 33ID at the Advanced Photon Source was supported by DOE-SC-BES under Contract DE-AC02-06CH11357 to UChicago Argonne, LLC as operator of Argonne National Laboratory. This research used computational resources of the National Energy Research Scientific Computing Center, which is supported by DOE-SC under Contract No. DE-AC02-05CH11231, and the Palmetto cluster at Clemson University.

[#]Present Address: Department of Chemistry; University of Puerto Rico, Río Piedras Campus; San Juan, PR 00931